# Evidence of Deterministic Components in the Apparent Randomness of GRBs: Clues of a Chaotic Dynamic

G. Greco[1*], R. Rosa[2], G. Beskin[3], S. Karpov[3], L. Romano[2], A. Guarnieri[4], C. Bartolini[4], and R. Bedogni[1]

[1] INAF-Astronomical Observatory of Bologna, Italy

[2] Department of Statistical Science, University of Bologna, Italy

[3] Special Astrophysical Observatory of Russia Academy of Science, Russia

[4] Department of Astronomy, University of Bologna, Italy

*Correspondence and requests for materials should be addressed to G.G. (giuseppe.greco8@unibo.it)

**Prompt γ-ray emissions from gamma-ray bursts (GRBs) exhibit a vast range of extremely complex temporal structures with a typical variability time-scale significantly short - as fast as milliseconds. This work aims to investigate the apparent randomness of the GRB time profiles making extensive use of nonlinear techniques combining the advanced spectral method of the Singular Spectrum Analysis (SSA) with the classical tools provided by the Chaos Theory. Despite their morphological complexity, we detect evidence of a non stochastic short-term variability during the overall burst duration - seemingly consistent with a chaotic behavior. The phase space portrait of such variability shows the existence of a well-defined strange attractor underlying the erratic prompt emission structures. This scenario can shed new light on the ultra-relativistic processes believed to take place in GRB explosions and usually associated with the birth of a fast-spinning magnetar or accretion of matter onto a newly formed black hole.**

Gamma-ray bursts (GRBs) are the most instantaneously powerful cosmic explosions known in the universe since the Big Bang. They are identified as brief, intense, and completely unpredictable flashes of high energy γ-rays in the sky. The prompt γ-ray emissions from GRBs exhibit a vast range of extremely complex temporal behaviors and any morphological classification scheme appears to be neither comprehensive nor systematic. Moreover, GRB profiles tend to be highly variable, showing flux variations of up to 100% on a time scale considerably shorter than the overall duration of the burst. The analysis of such variability has characterized much of the scientific work on GRBs[1]. In fact, it is crucial to understand the physical mechanisms driving the internal engine of GRBs, which remains hidden from direct observation. We demonstrate, for the first time, that the prompt temporal variability of GRBs does not follow a pure random behavior as previously assumed, but instead responds to a low-dimension deterministic dynamic which appears to govern the overall burst duration. A universal *strange attractor* - underlying the erratic prompt structure - suggests that a coherent deterministic mechanism is taking place during the entire dynamical evolution of a GRB explosion. Moreover, we detect clues of a chaotic dynamic that offers the first observational evidence of *deterministic chaos* in relativistic astrophysical phenomena placed at cosmological distance. A deterministic chaotic system is characterized by a high sensitivity to initial conditions. By exhibiting an apparently stochastic behavior, such sensitivity masks the deterministic laws of their intrinsic physical processes. For the very first time, this particular type of dynamics was observed in objects capable of producing significant space-time distortion, even possibly emitting gravitational waves[2].

Results

**Data Sample**. To make our analysis meaningful, we collected a very high-quality data sets with more than 3000 points for each temporal sequence sampled with a homogeneous time bin-width of 64ms. A Signal-to-Noise level (S/N) above 50 is required to ensure the reliability of the prompt structures during the transient evolution of a GRB. The list of prompt time profiles was taken from the daily updated compilation of Swift team. Throughout the January 2005 to September 2010 period, two events match our selection criteria, namely GRB 051117 and GRB 100814. In an attempt to remain brief, this paper includes the graphs of GRB 050711 exclusively. The plots obtained for GRB 100814 are largely similar.

**Search for Deterministic Components: Monte Carlo SSA.** The Singular Spectrum Analysis[3,4] (SSA) belongs to the class of methods which use orthogonal functions, allegedly more efficient because calculated on the basis of data rather than on the basis of a fixed selected base as in Fourier and Wavelet Transforms. Indeed, this property makes the SSA a very effective statistical method in the decomposition of the original time series into a defined number of independent and interpretable components, e.g. low-frequency trend, anharmonic oscillatory components, periodicities with varying amplitudes, or structureless noise. It permits to separate signal from noise even if the Signal-to-Noise level varies when recording the original time series - a typical feature of GRBs' prompt emissions. At all time in this research project, a Monte Carlo SSA test[5,6,7] was used to distinguish between genuine deterministic components and components generated by a pure red-noise process. Red-noise is known to be significantly relevant in several natural systems. It is dominated by cycles of low frequency in its power spectrum and exhibits significant autocorrelations that decay over time. When dealing with red-noise, a first-order autoregressive process is usually considered: *AR(1)*, given by $x_t=\phi x_{t-1}+\varepsilon_t$ with $0<\phi<1$ and $\varepsilon_t$ independent identically distributed normal errors. In this case, a total of 10000 surrogate realizations were used for the computation of percentiles for each eigenvalue $\lambda_L$. Using the Monte Carlo SSA test, we estimated the parameters of the *AR(1)* model starting from the very same time series, and using a maximum-likelihood criterion. For instance, if an eigenvalue $\lambda_L$ lies outside a 99% noise percentile, then the red-noise null hypothesis can be rejected with this confidence. Otherwise, that particular SSA component of the time series cannot be considered as significantly different from red-noise.

The shape of the so-called SSA Eingespectrum is analyzed to identify possible evidence of deterministic activity in the prompt emission from GRBs selected in our sample. The SSA Eingespectrum plot shows the eigenvalue, $\lambda_L$, ranked by order that provides the variance of the time series in the direction specified by the corresponding eigenvector ***E**$_L$*. Each eigenvalue represents the fraction of total variance explained by the associated component. More specifically, if two components explain more or less the same variance and their modes are in phase quadrature, they may represent an oscillatory patterns[8]. Alternatively, by selecting low-frequency SSA components, we can identify nonlinear slow trends. In this work, Kendall's $\tau$ nonparametric trend tests are performed to identify those components that are significantly non-stationary over the length of the time series at 99% confidence levels (Fig. 1). Once the eigenvalues identifying the trend have been determined, the de-trended prompt light curve is to be tested against the null-hypothesis, $H_0$: the variability nature of the data are consistent with a pure *AR(1)* random noise. We found that the first eigenvalues are dominant and lie outside the intervals that define a purely stochastic behavior. Furthermore, we were able to reject the null hypothesis at the 0.01 level, as the variances showed to be significantly different from the noise-variance. The results of this analysis are showed in Fig. 2. In Fig. 3 can be observed the detailed steps of the SSA reconstruction process. As evidenced, a deterministic signal clearly emerges out of a significant fraction of random walk noise, from the time of the original signal to the end of this analysis. The figure depicts the de-trend and de-noise version of the original time series in which the red-noise and the trend component have been isolate.

A detailed analysis of background region pre- and post- GRB explosion confirms the validity of the results exposed in this paper. We were indeed able to demonstrate that the background perfectly follows a red-noise behavior, with no exceptions. Fig. 4 shows the SSA Eigenspectrum of the local-background region of GRB 050711.

**Phase Space Reconstruction and Correlation Dimension**. Based on time-delay embedding theorem of Takens[9], we reconstructed the phase space of the dynamical system from the time-sequence of observations collected through our experiments. The reconstruction preserves the essential properties of the dynamic system (topological structure of the attractor like the correlation dimension and Lyapunov exponents), but it does not preserve the geometric shape of structures in phase space. To reconstruct the state space, we use the method of delay (MOD). With the MOD, the spatial structure of the dynamical system is visualized simply by plotting the scalar quantity of the signal sampling at regular time intervals, $\tau_s$, against the time-delayed values of the scalar measurements $\tau$ . Thus, we can build a vector in $m$ dimensions (generally referred to as embedding dimension) in order to examine the dynamics of the signal. Once a proper time delay $\tau$ and embedding dimension $m$ have been found, a phase space for the dynamic analysis is provided. There is a large literature on the optimal choice of the embedding parameters $m$ and $\tau$. The usual autocorrelation function and the time-delayed mutual information provide important suggestions about reasonable delay times while the false neighbors statistic can give guidance about the proper embedding dimension. The study of trajectories in phase space is essential to discriminate chaos from pure randomness, or more generally a deterministic system from a stochastic dynamic. To this aim, it is important to measure the geometrical structure of the attractor, so-called correlation dimension $D_2$. The estimation of $D_2$ gives the minimum number of variables necessary to describe the state of the system, at any given time. A non-integer result for the correlation dimension indicates the fractality of the system. We followed the method proposed by Grassberger & Procaccia[10] according to which the saturation value of $D_2$ gives the attractor dimension. For a truly random signal, the correlation dimension graph does not display any kind of saturation, indicating a high-dimensional random behaviour with $D_2 \rightarrow \infty$. Inherent deterministic signals, on the other hand, have a distinct spatial structure, meaning that their correlation dimension will saturate at some point, as embedding dimension $m$ is increased.

A strange attractor can be chaotic or not, depending on cases. By *strange* we refer to metric properties such as fractal dimensions, while *chaotic* reflects dynamic properties including the exponential divergence of nearby trajectories in phase space (sensitivity to initial conditions).

**Theory of Chaos: Maximum Lyapunov Exponent**. The Theory of Chaos (or Deterministic Chaos) deals with the study of systems that exhibit a complex temporal evolution. Such theory, indeed, researches whether the said systems can be described in their whole dynamic evolution using only a relatively small number of differential equations. It is to be highlighted that their apparent randomness is not due to external perturbations - inherently uncontrollable - as in the case of molecules' impact on motes; nor is it due to the physical principle that characterizes the microscopic world - which prevents the prediction of the exact point of arrival when measuring quantum particles (Heisenbergs Uncertainty Principle). Rather, it springs from the same nonlinear deterministic law which governs the few individuals components of the system. An important feature of chaotic systems is the exponential growth of the distance between trajectories initially very close (sensitivity to initial conditions). The parameter used to measure such growth rate is the maximum Lyapunov exponent $\lambda_{max}$. A positive value of $\lambda_{max}$ indicates the sensitivity to initial conditions, typical hallmark of chaos.

Further investigations on the nature of the variability of the above described components were conducted measuring the dynamical invariant of their attractors. The phase space portrait of the deterministic component discovered in GRB 050711 is plotted in Fig. 5. Considering a small radius ε and an embedding dimension ranging from 6 to 11, the obtained correlation dimension shows evidence of a fractal nature with $D_2 \approx 2.4$ and $D_2 \approx 2.8$ for GRB 050711 and for GRB 100814, respectively. Because of the small number of points collected in these time series, the previous analysis provided a lower limit than the accepted correlation dimension. Furthering our analysis, we set to measure the maximum Lyapunov exponent using the algorithm of Kantz[11]. In addition, we used the method explained in Giannerini & Rosa[12] to assess standard errors and confidence intervals for the estimated maximal Lyapunov exponent. As a result, we obtained $\lambda_{max} = 0.0089 \pm 0.0007$ and $\lambda_{max} = 0.0102 \pm 0.0005$ for the strange attractor of GRB 050711 and GRB 100814, respectively.

**Additional Evidence of Chaotic Dynamic: Recurrence Plot.** Finally, we were able to visualize the GRB dynamic via the recurrence plot analysis[13]. Such analysis provides a graphical representation of the patterns in a time series and was first introduced to visualize the time-depended behaviour of the dynamical systems. It represents the recurrence of the phase space trajectory at a certain state, exhibiting characteristic large scale and small-scale patterns that are caused by the particular dynamic system under examination. More specifically, the presence of short diagonal segments indicates the presence of the so-called unstable periodic orbits embedded in the chaotic attractor of a dynamic system. By visual comparison, we note that the structures of GRB recurrence plots have similarities with those of know chaotic attractors, increasing the evidence that a chaotic dynamic is possible and not entirely negligible (Fig. 6).

Discussion

The use of the advanced spectral method of the SSA, together with the classical tools provided for in the Theory of Chaos, proved largely successful in the analysis of the complex morphological structure of the prompt emissions from GRBs. It results from such analysis that: *(i)* the prompt emission of GRBs can be described by a low-frequency trend component that represents more than 90% of the total variance of the time profile. To this trend are superimposed deterministic oscillations with lower variance. A significant fraction of red-noise affects the entire dynamic evolution. *(ii)* Despite the extremely complicated and random profile of the prompt γ-ray signals, we found a well-defined strange attractor. This implies that the nature of the prompt time variability is not purely stochastic. Rather, similar inherent physical processes appear to take place during the entire dynamic evolution of GRB explosions. *(iii)* Evidence was found of a fractal ($D_2 \approx 2.4$ and $D_2 \approx 2.8$) and chaotic nature ($\lambda_{max}$ = 0.0089±0.0007 and $\lambda_{max}$ =0.0102±0.0005) of the GRB attractors.

The presence of a low-dimension chaotic dynamics allow us to simplify the descriptive theory of GRB prompt mechanisms, as well as, to test and constrain the different theoretical scenarios. In fact, the Theory of Chaos deals with the study of systems that exhibit a complex temporal evolution. Such theory, indeed, researches whether the said systems can be described in their whole dynamic evolution using only a relatively small number of differential equations. The estimation of correlation dimension $D_2$ gives the minimum number of variables necessary to describe the state of the system, at any given time, making the GRBs signal more *tractable* from a mathematical point of view.

As a conclusion, we can safely suggest that such dynamic results common to various astrophysical objects detected in our near to distant Universe, as also confirmed by the evidence of a deterministic chaos gathered within the Solar System itself[14]. Continuous improvements of sampling techniques in the field of astrophysics provide for the possibility of many future discoveries regarding chaotic behaviour in cosmological environments.

Methods

Formally, SSA method can be summarized in two main stages, each stage consisting of two principal steps (*i*) decomposition stage: embedding and singular value decomposition, and (*ii*) reconstruction stage: grouping and diagonal averaging. In the first step, an one-dimensional time series $F = (f_0, f_1,..., f_{N-1})$ of length $N$ is recast as an $L$-dimensional time series forming the so-called trajectory matrix **X** of the system. We emphasize that the trajectory matrix **X** is obtained by setting a priori a specified window length $L$. The second step is the singular value decomposition (SVD) of the matrix X, which can be obtained via eigenvalues $(\lambda_1,....,\lambda_L)$ and eigenvectors $(\mathbf{E}_1,...,\mathbf{E}_L)$ of the matrix $\mathbf{S} = \mathbf{XX}^T$ of size $L \times L$. In the first and second steps of the second stage, i.e. the "reconstruction stage", the components are grouped together. Each matrix of the grouped decomposition is then used to reconstruct a specific component of the original time series with the same length $N$ than the starting signal.

In conducting the SSA tests, a suitable window length $L$ must be chosen. $L$ determines the number of lagged vectors that are used to form the trajectory matrix, and thus, the resolution of the decomposition. Following the norm, $L$ was varied and tested over a large range of values before electing a window length of $L \sim N/15$.

Meanwhile, the attractor reconstruction and its invariant quantities ($D_2$, $\lambda_{max}$) were obtained by setting embedding parameters $m = 3$ and $\tau = 25$ s as suggested by the false neighbor statistics and the mutual information plot, respectively.

The list of prompt GRB profiles was taken from the daily updated compilation of Swift/BAT team *http://www.nasa.gov/mission_pages/swift/team/index.html*.

References

1. Piran, T. The physics of gamma-ray bursts. *Reviews of Modern Physics*. **76**, 1143-1210 (2004).
2. Corsi, A. Gravitational Waves and High energy emission from GRBs: an observational Review. *38th COSPAR Scientific Assembly*. Held 18-15 July 2010, in Bremen, Germany. **16**, E16-0002-10 (2010).
3. Elsner, J. B. and Tsonis, A. A. Singular Spectrum Analysis: A New Tool in Time Series Analysis. *Plenum Press*, New York and London (1996).
4. Golyandina, N., Nekrutkin, V. and Zhigljavsky, V. Analysis of Time Series Structure: SSA and Related Techniques. *Monographs on Statistics and Applied Probability 90. Chapman & Hall/CRC*, Boca Raton (2001).
5. Vautard, R., Yiou, P. and Ghil, M. Singular-spectrum analysis: A toolkit for short, noisy chaotic signals. *Physica D*. **58**, 95-126 (1992).
6. Allen, M. R. and Smith, L. A. Monte Carlo SSA: detecting irregular oscillations in the presence of coloured noise. *J. Climate*. **9**, 3373-3404 (1996).
7. Ghil, M. *et al.* Advanced spectral methods for climatic time series. *Rev. Geophys*. **40**, 1-41 (2002).
8. Ghil, M. and Mo, K. Intraseasonal oscillations in the global atmosphere-Part I: Northern Hemisphere and tropics. *J. Atmos. Sci*. **48**, 752-779 (1991).
9. Takens, F. Detecting strange attractors in turbulence. *Springer Lecture Notes in Mathematics*. **898**, 366-381 (1981).
10. Grassberger, P. and Procaccia, I. Measuring the Strangeness of Strange Attractors. *Physica D: Nonlinear Phenomena*. **9**, 189-208 (1982).
11. Hegger, R., Kantz, H. and Schreiber, T. Practical implementation of nonlinear time series methods: The TISEAN package. *Chaos*. **413**, 10.1063-1.166424 (1999).
12. Giannerini, S. and Rosa, R. New resampling method to assess the accuracy of the maximal Lyapunov exponent estimation. *Physica D*. **155**, 101-111 (2001).
13. Marwan, N., Romano, M. C., Thiel, M., Kurths, J. Recurrence Plots for the Analysis of Complex Systems, *Physics Reports*, **438**, 237-329 (2007).
14. Murray, N. and Holman, M. The Origin of Chaos in the Outer Solar System. *Science*. **283**, 1877-1879 (1999).

Acknowledgments

We acknowledge Taka Sakamoto and Scott D. Barthelmy for Swift/BAT data used in this work. G. G. acknowledges the *Fondazione Cassa di Risparmio in Bologna* for INAF-fellowship and M. Nardi for helpful suggestions.

Author contribution statement

GG, RR and GB wrote the main manuscript text and GG, LR, AG, CB, RB, SK prepared figures 1-6. All authors reviewed the manuscript.

Additional information.

The author declares no competing financial interests

Figure captions

Fig.1. SSA Eigenspectrum of prompt emission from GRB 050117. Kendall's $\tau$ trend tests are performed to identify those components that are significantly non-stationary over the length of the time series at 99% confidence levels. These eigenvalues are circled in red.

Fig.2. Monte Carlo SSA Eigenspectrum of prompt emission from GRB050117 in which the low-frequency trend is subtracted. The error bars represent the interval between the 0.01$^{th}$ and 99.9$^{th}$ percentiles, and eigenvalues that lie outside this range are significantly different from those generated by a red-noise *AR*(1) process against which they are tested by using 10000 Monte Carlo simulation.

Fig.3. SSA Reconstruction. The original time series is *de-trended* and *de-noised* in according to the SSA Eigenspectrum. As evidenced, a deterministic signal clearly emerges out of a significant fraction of random walk noise, from the time of the original signal to the end of this analysis.

Fig.4. Monte Carlo SSA Eigenspectrum test of the local-background region of GRB 050117. The error bars represent the interval between the 0.01$^{th}$ and 99.9$^{th}$ percentiles, and eigenvalues that lie outside this range are significantly different from those generated by a red-noise *AR*(1) process against which they are tested by using 10000 Monte Carlo simulation.

Fig.5. Phase Space Portrait of the deterministic components discovered in the prompt emission from GRB 050117. The three axes represent the delay coordinate vector with τ = 25 s.

Fig.6. Recurrence Plot of the deterministic components discovered in the prompt emission from GRB 050117.

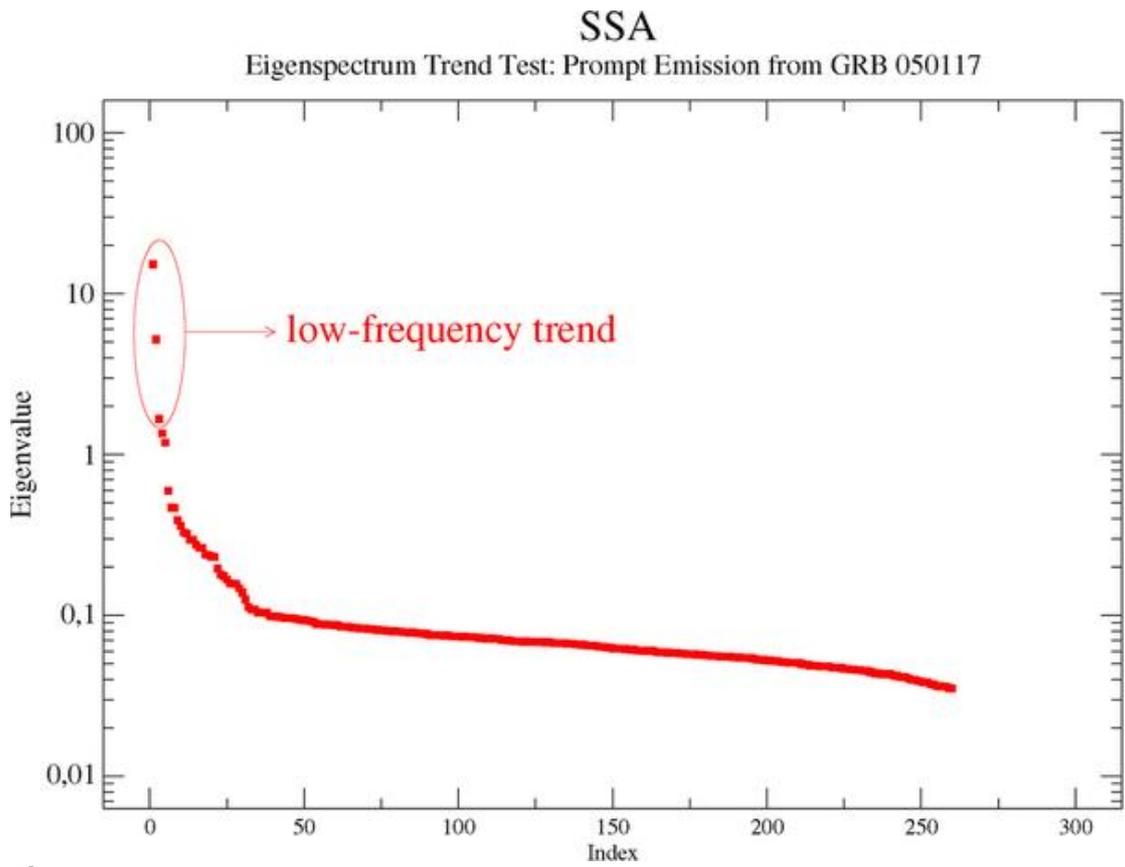

**Fig.1**

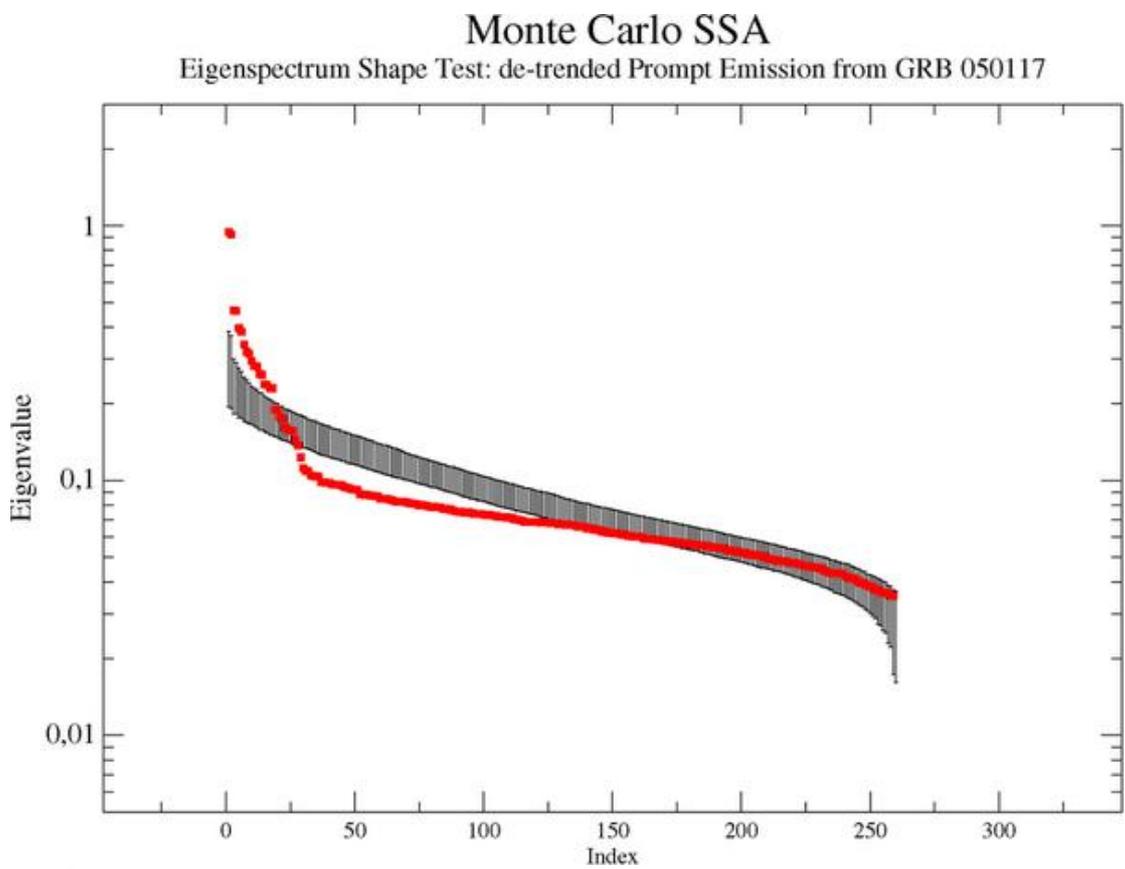

**Fig.2**

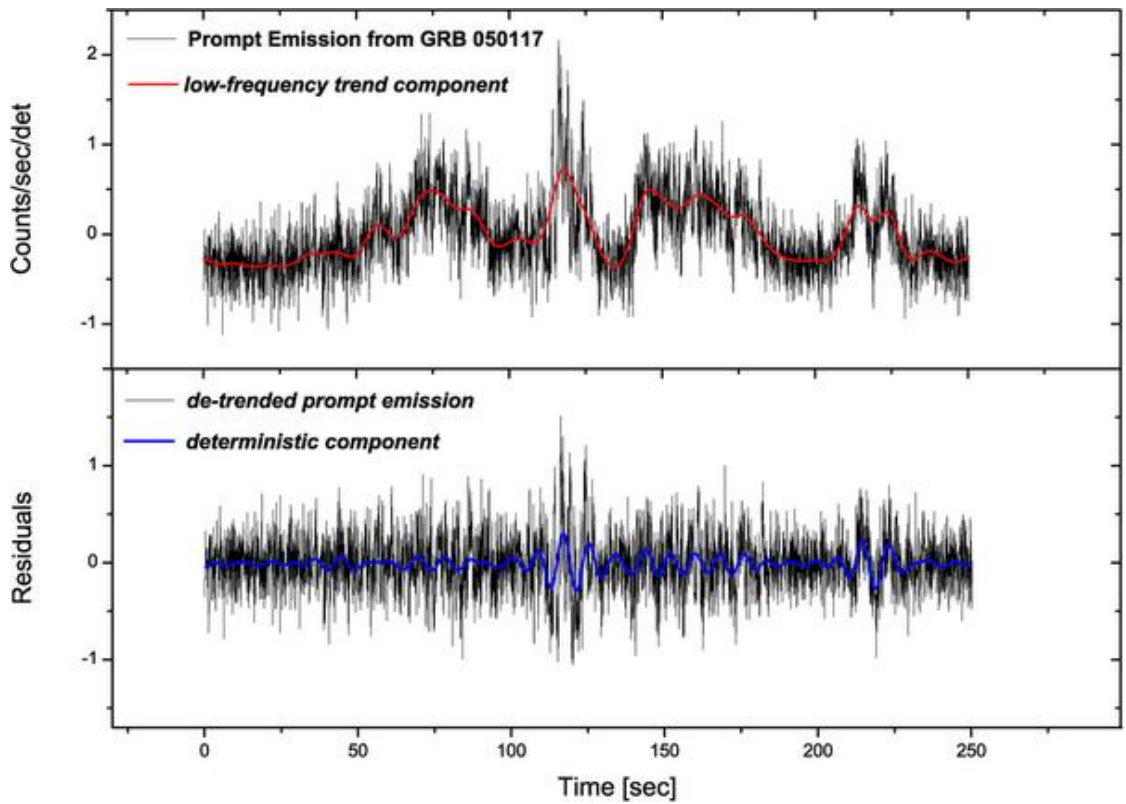

**Fig.3**

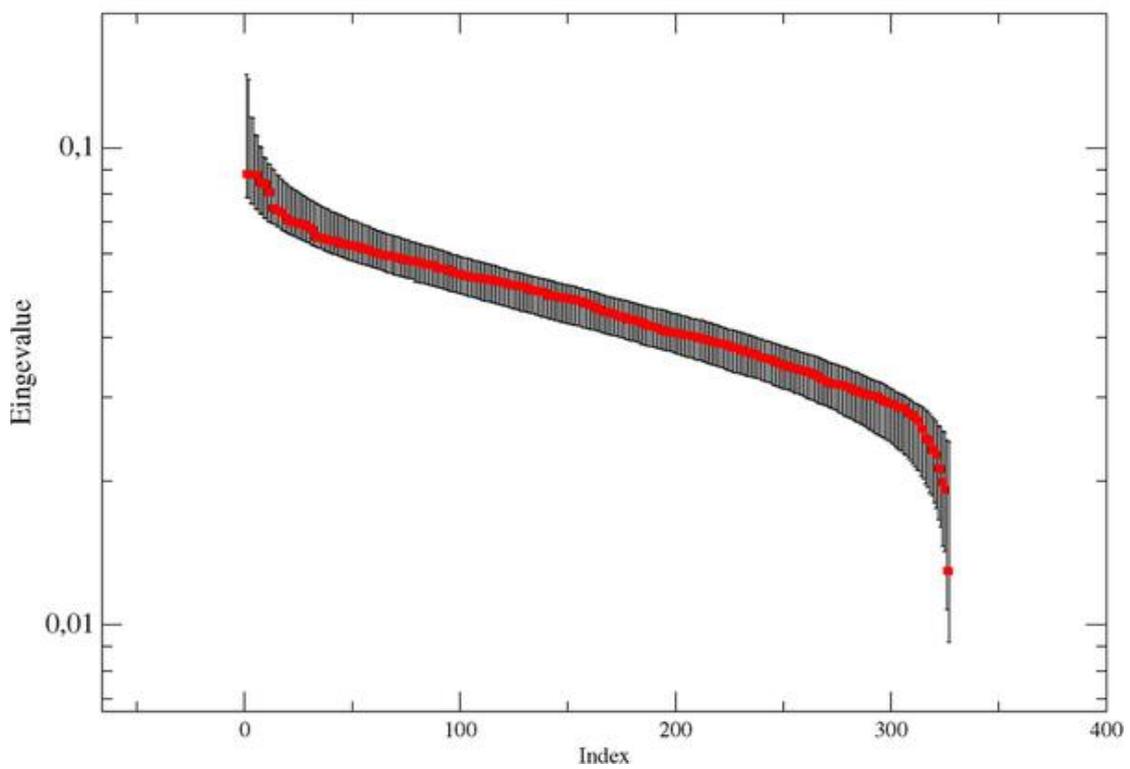

**Fig.4**

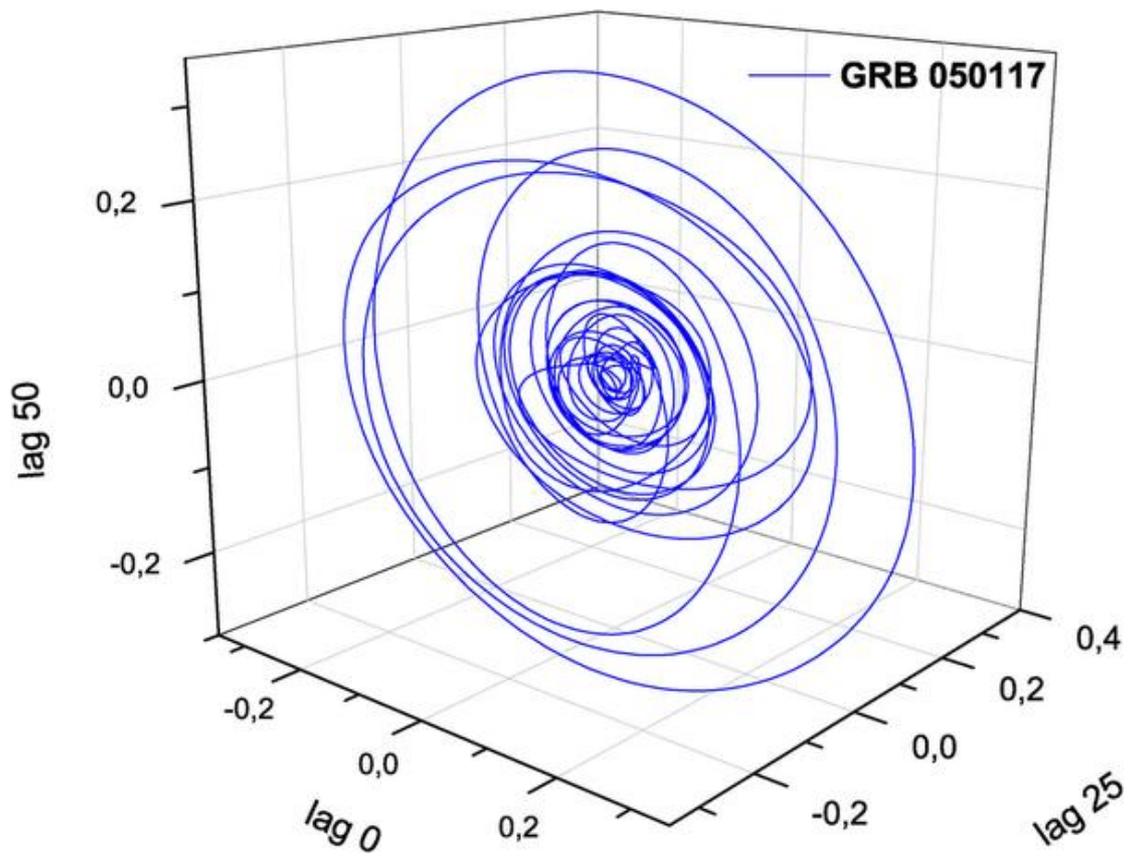

**Fig.5**

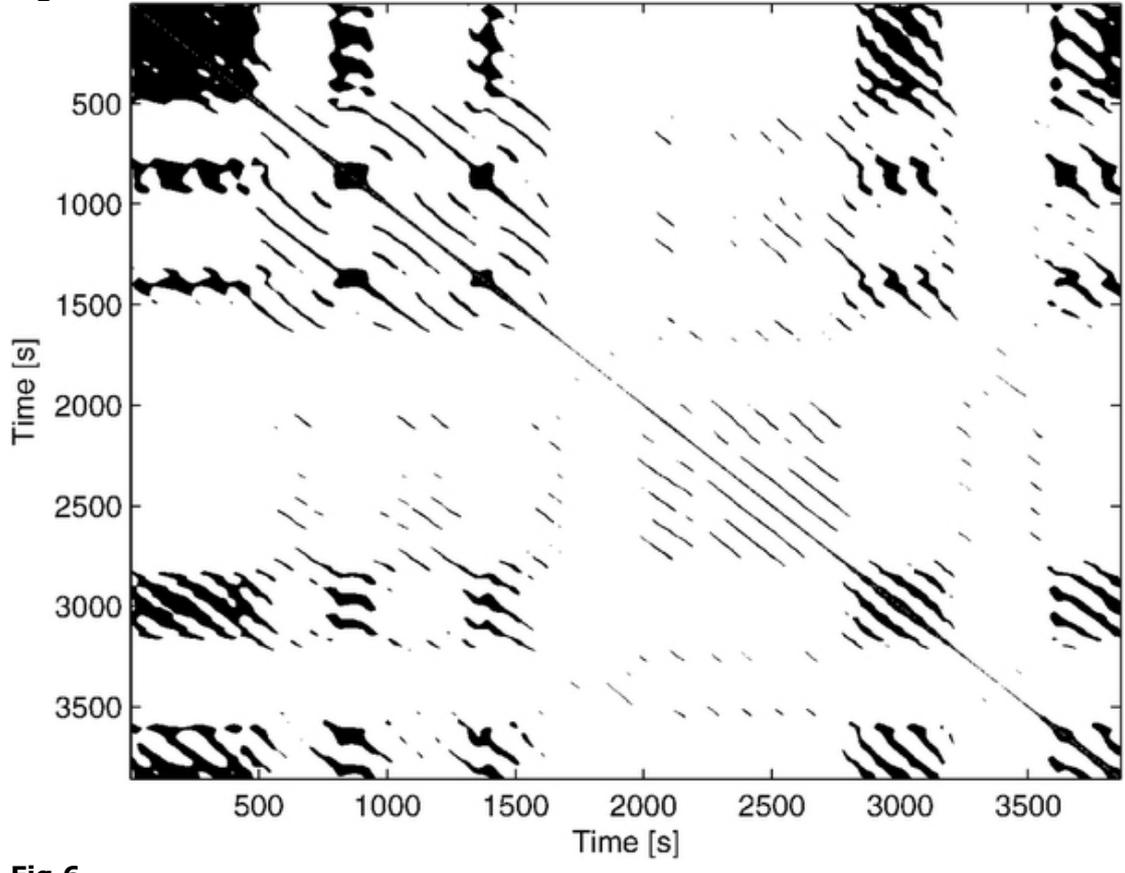

**Fig.6**